\documentclass[proceedings]{JHEP3}
\usepackage{eurosym}
\usepackage{amssymb}
\usepackage{amsfonts}
\usepackage{amsmath,epsfig}
\usepackage{graphicx}

\newbox\mybox

\newcommand\fverb{\setbox\mybox=\hbox\bgroup\verb}
\newcommand\fverbdo{\egroup\medskip\noindent\fbox{\unhbox\mybox}\ }
\newcommand\fverbit{\egroup\item[\fbox{\unhbox\mybox}]}
\conference{Pseudo-Hermitian approach to GT in non-Abelian non-Hermitian QFT}
\abstract{We generalise previous studies on the extension of Goldstone's theorem
from Hermitian to non-Hermitian quantum field theories with Abelian symmetries to theories possessing a glocal non-Abelian symmetry. 
We present a detailed analysis for a non-Hermitian field theory with two complex two component scalar fields possessing a SU(2)-symmetry and indicate how
our finding extend to the general case. In the PT-symmetric regime and at the standard exceptional point the Goldstone theorem is shown to apply, 
although different identification procedures need to be employed. At the zero exceptional points 
the Goldstone boson can not be identified. Comparing our approach, based on the pseudo-Hermiticity of the model, to an alternative approach that utilises 
surface terms to achieve compatibility for the non-Hermitian system, we find that the explicit forms of the Goldstone boson fields are different.}

\title{Pseudo-Hermitian approach to Goldstone's theorem in non-Abelian
non-Hermitian quantum field theories}
\author{Andreas Fring and Takanobu Taira \\
Department of Mathematics, City, University of London,\\
Northampton Square, London EC1V 0HB, UK \\
E-mail: a.fring@city.ac.uk, takanobu.taira@city.ac.uk}

\input{tcilatex}
\begin{document}

\section{Introduction}

The extension from quantum field theories with Hermitian actions to those
with a non-Hermitian actions has been addressed recently for various
concrete systems, such as scalar field theory with imaginary cubic
self-interaction terms \cite{benderphi31,shalabyphi31}, field theoretical
analogues to the deformed harmonic oscillator \cite{bender2018p},
non-Hermitian versions with a field theoretic Yukawa interaction \cite%
{alexandre2015non,rochev2015hermitian,korchin2016Yuk,laureyukawa}, free
fermion theory with a $\gamma _{5}$-mass term and the massive Thirring model 
\cite{bender2005dual}, $\mathcal{PT}$-symmetric versions of quantum
electrodynamics \cite{bender1999nonunitary,milton2013pt} and $\mathcal{PT}$%
-symmetric quantum field theories in higher dimensions \cite{benderH2018p}.

The generalisations also include Goldstone's theorem \cite%
{nambu1961dynamical,goldstone1961field} and the Higgs mechanism \cite%
{englert,higgs,kibble} \cite%
{alexandre2018spontaneous,mannheim2018goldstone,fring2019goldstone,alex2019}%
. Both of these mechanisms are governed by the continuous symmetries of the
theories, global or local, respectively, that might by spontaneously broken
by some vacuum states. The special feature of non-Hermitian systems is that
an additional discrete antilinear symmetry \cite{EW} is superimposed on top
of the continuous symmetries, that can also be spontaneously broken, albeit
not exclusively for the ground state in this case. The regime in which the
discrete symmetry is broken is regarded as unphysical. In general, the\
antilinear symmetry separates the parameter space of the theory into regimes
of different types of behaviour. The physical subspace is bounded by the
values for which the eigenvalues of the mass squared matrix acquire an
exceptional point, a singularity or become zero. It is the interplay between
these two types of symmetries, continuous and discrete, that produce very
interesting and novel behaviour when compared to the standard Hermitian
setting.

There is a well known problem that seems to suggest that non-Hermitian
quantum field theories are inconsistent, see e.g. \cite%
{alexandrefoldy,alexandre2017symmetries,bender2005dual}. However, just as
for non-Hermitian quantum theories \cite{Geyer,Bender:1998ke,Alirev,PTbook}
there are methods and techniques to overcome these issues to obtain a
perfectly consistent theory. The conundrum for the quantum fields theories
consists of the feature that the two sets of equations of motion, derived
from functionally varying the action with respect to the scalar fields on
one hand and with respect to their complex conjugates on the other, are
incompatible. So far two distinct alternative propositions have been made to
overcome this issue. Alexandre, Ellis, Millington and Seynaeve proposed to
apply a non-standard variational principle by keeping some non-vanishing
surface terms \cite{alexandre2018spontaneous,alex2019} or, in line with the
pseudo-Hermitian/$\mathcal{PT}$-symmetric quantum mechanical approach \cite%
{Geyer,Bender:1998ke,Alirev,PTbook}, one may seek a consistent equivalent
similarity transformed Hermitian action, as pursued by Mannheim and the
present authors \cite{mannheim2018goldstone,fring2019goldstone}. While some
features are the same in both approaches, e.g. both versions predict the
same number of massless Goldstone bosons that is expected from Goldstone's
theorem, they also differ in several aspects. While in the former
proposition Noether's theorem is evaded the latter is based on the standard
variational principle leading to standard Noether currents. Moreover with
regard to the Higgs mechanism the "\textit{surface term approach}" predicts
that the gauge particle becomes massive in the local case \cite{alex2019},
whereas the "\textit{pseudo-Hermitian approach}" leads to a theory in which
the gauge particle remains massless at the exceptional point \cite%
{mannheim2018goldstone}. Here we also find that the explicit form of the
Goldstone bosons differs.

Previous considerations were focused on the analysis of non-Hermitian
systems with a global and local Abelian $U(1)$-symmetry, they were recently
extended to non-Abelian theories within the \textit{surface term approach }%
\cite{alex2019}. Here we also extend these studies to the non-Abelian case
by applying the \textit{pseudo-Hermitian approach}. We analyse in detail a
non-Hermitian scalar field theory with two complex two component scalar
fields possessing a $SU(2)$-symmetry and an overall discrete antilinear
symmetry. We compare our results to those obtained in \cite{alex2019} by
means of the surface term approach.

In section 2 we discuss the generalities of the pseudo-Hermitian approach to
achieve compatibility in non-Hermitian quantum field theories, with an
emphasis on how it modifies the identification of the mass squared matrix
and Goldstone's theorem. In section 3 we discuss a concrete model with two
complex scalar fields in the fundamental representation, by deriving an
equivalent Hermitian action for the model, discussing its $SU(2)$-symmetry,
its vacua, mass squared matrices, physical regions and identifications of
the Goldstone bosons in the different regimes. We state our conclusions in
section 4.

\section{Pseudo-Hermitian approach to spontaneously broken symmetries}

We consider here complex scalar quantum field theories described by actions
of the following generic type%
\begin{equation}
\mathcal{I}=\int d^{4}x\left[ \partial _{\mu }\phi \partial ^{\mu }\phi
^{\ast }-V(\phi )\right] ,  \label{1}
\end{equation}%
with $n$-component complex scalar fields $\phi =(\phi _{1},\ldots ,\phi
_{n}) $ and potential $V(\phi )$. The action is assumed to possess three
general properties: i) It is invariant under a global continuous symmetry $%
\phi \rightarrow \phi +\delta \phi $ with $V(\phi )=V(\phi +\delta \phi )$.
The symmetry is, for instance, generated by a Lie group \textbf{g} with Lie
algebraic generators $T$, so that being global implies an infinitessimal
change $\delta \phi =\alpha T\phi $ with $\alpha $ being a small parameter
and $\partial _{\mu }(\alpha T)=0$. ii) It is invariant under a discrete
antilinear symmetry $\phi (x_{\mu })\rightarrow U\phi ^{\ast }(-x_{\mu })$,
with $U$ being a constant unitary matrix. These symmetries may be viewed as
modified $\mathcal{CPT}$-symmetries. When $U\rightarrow \mathbb{I}$ the
symmetry reduces to the standard $\mathcal{CPT}$-symmetry. iii) The
potential $V(\phi )$ is not Hermitian, that is $V(\phi )\neq V^{\dagger
}(\phi )$.

At first sight such type of theories appear to be inconsistent as the two
sets of equations of motion obtained by functionally varying the action $%
\mathcal{I}$ separately with respect to the fields $\phi _{i}$ and $\phi
_{i}^{\ast }$, $\delta \mathcal{I}_{n}/\delta \phi _{i}=0$ and $\delta 
\mathcal{I}_{n}/\delta \phi _{i}^{\ast }=0$, are in general incompatible
when $U\neq \mathbb{I}$. One may, however, overcome this problem by using a
non-standard variational principle combined with keeping some non-vanishing
surface terms \cite{alexandre2018spontaneous,alex2019} or alternatively by
exploiting the fact that the content of the theory is unaltered as long as
the equal time commutation relations are preserved and carry out a
similarity transformation that guarantees that feature \cite%
{bender2005dual,mannheim2018goldstone,fring2019goldstone}. Hence, in the
latter approach one seeks a Dyson map $\eta $, named this way in analogy to
its quantum mechanical counterpart \cite{Dyson}, to construct a new
equivalent action 
\begin{equation}
\hat{{\mathcal{I}}}=\eta \mathcal{I}\eta ^{-1}=\int d^{4}x\left[ \partial
_{\mu }\phi \hat{I}\partial ^{\mu }\phi ^{\ast }-\hat{V}(\phi )\right] ,
\end{equation}%
with the difference that now the transformed potential is Hermitian $\hat{V}%
(\phi )=\hat{V}^{\dagger }(\phi )$. The matrix $\hat{I}$ is a result of the
similarity transformation.

Next it is in general useful to convert the complex scalar field theory into
one involving only real valued fields by decomposing the $n$ complex scalar
fields into real and imaginary parts as $\phi =1/\sqrt{2}(\varphi +i\chi )$
with $\varphi $,$\chi \in \mathbb{R}$. Defining then a real $2n$-component
field $\Phi =(\varphi _{1},\ldots ,\varphi _{n},\chi _{1},\ldots ,\chi _{n})$%
, possibly with the fields in different order to block diagonalize the mass
squared matrix, the new action $\hat{{\mathcal{I}}}$ may be re-written as%
\begin{equation}
\hat{{\mathcal{I}}}=\int d^{4}x\left[ \frac{1}{2}\partial _{\mu }\Phi
I\partial ^{\mu }\Phi ^{\ast }-\hat{V}(\Phi )\right] .
\end{equation}%
Analyzing the action in this form, the extension of Goldstone's theorem from
the Hermitian to the non-Hermitian case is easily established. At first we
identify various types of vacua $\Phi _{0}$ by solving 
\begin{equation}
\left. \frac{\partial \hat{V}(\Phi )}{\partial \Phi }\right\vert _{\Phi
=\Phi _{0}}=0.  \label{vac}
\end{equation}%
The continuous global symmetry $\Phi \rightarrow \Phi +\delta \Phi $, i.e. $%
\hat{V}(\Phi )=\hat{V}(\Phi +\delta \Phi )=$ $\hat{V}(\Phi )+\nabla \hat{V}%
\left( \Phi \right) ^{T}\delta \Phi $, then implies%
\begin{equation}
\frac{\partial \hat{V}(\Phi )}{\partial \Phi _{i}}\delta \Phi _{i}(\Phi )=0.
\end{equation}%
Differentiating this equation with respect to $\Phi _{j}$ and evaluating the
result at a vacuum $\Phi _{0}$, determined by (\ref{vac}), yields%
\begin{equation}
\left. \frac{\partial ^{2}\hat{V}(\Phi )}{\partial \Phi _{j}\partial \Phi
_{i}}\right\vert _{\Phi =\Phi _{0}}\delta \Phi _{i}(\Phi _{0})+\left. \frac{%
\partial \hat{V}(\Phi )}{\partial \Phi _{i}}\right\vert _{\Phi =\Phi
_{0}}\left. \frac{\partial \delta \Phi _{i}(\Phi )}{\partial \Phi _{j}}%
\right\vert _{\Phi =\Phi _{0}}=0.  \label{con}
\end{equation}%
Since the last term vanishes, due to (\ref{vac}), we are left with two
options to solve (\ref{con}). Either the vacuum is left invariant such that $%
\delta \Phi _{i}(\Phi _{0})=0$ or the vacuum breaks the global symmetry and $%
\delta \Phi _{i}(\Phi _{0})\neq 0$. Denoting $\theta _{0}:=\delta \Phi
_{i}(\Phi _{0})$ and multiplying (\ref{con}) by $\hat{I}$ we obtain 
\begin{equation}
\hat{I}H(\Phi _{0})\theta _{0}=M^{2}\theta _{0}=0,  \label{Gold}
\end{equation}%
where $H(\Phi _{0})$ is the Hessian of the potential $\hat{V}(\Phi )$
evaluated at the vacuum $\Phi _{0}$ and $M^{2}$ is the mass squared matrix.
The occurrence of the matrix $\hat{I}$ results from the similarity
transformation and is therefore the trace of the feature that the potential
is non-Hermitian. It also has the effect that $M^{2}$ is no longer Hermitian
either. We can now read off Goldstone's theorem for non-Hermitian systems
from (\ref{Gold}). When the vacuum is left invariant by the global symmetry
transformation we have $\theta _{0}=0$ so that there is no restriction on $%
M^{2}$. However, when the vacuum breaks the global symmetry we have $\theta
_{0}\neq 0$ so that $\theta _{0}$ becomes a null vector for $M^{2}$. Thus,
in this case we have a zero mass particle, that is identified as a Goldstone
boson.

Assuming that the symmetry is generated by a Lie group \textbf{g}, we may
repeat this argument for each Lie algebraic generator $T$ so that we obtain
a Goldstone boson for each generator that when acting on the vacuum $\Phi
_{0}$ produces a different one. The crucial difference, when compared to the
scenario with Hermitian potentials, is that here $M^{2}$ is also not
Hermitian. This means that the physical regimes are determined by the
discrete antilinear symmetries. Referring to this symmetry as $\mathcal{PT}$%
-symmetry \cite{Bender:1998ke,PTbook} in a wider sense, we may encounter $%
\mathcal{PT}$-symmetric regimes with real mass spectra, exceptional points
with non-diagonalisable mass matrix, zero exceptional points, singularities
and a spontaneously broken $\mathcal{PT}$-symmetric regime with unphysical
complex conjugate masses. As shown in \cite{fring2019goldstone} the
identification of the Goldstone boson is different in these regimes and in
parts impossible.

Below we will also make use of the general property that the expansions
around two vacua, say $\phi _{0}^{1}$ and $\phi _{0}^{2}$, that are related
by the symmetry transformation $\mathcal{T}$ of the potential $V(\phi )=V(%
\mathcal{T}\phi )$ as $\mathcal{T}\phi _{0}^{1}=\phi _{0}^{2}$ with $%
\mathcal{T}^{T}=\mathcal{T}^{-1}$ yield to theories with mass squared matrix
possessing the same eigenvalues. This can be seen from 
\begin{eqnarray}
V(\phi +\phi _{0}^{1}) &=&V(\phi +\mathcal{T}^{-1}\phi _{0}^{2})=V(\mathcal{T%
}^{-1}(\mathcal{T}\phi +\phi _{0}^{2}))=V(\mathcal{T}\phi +\phi _{0}^{2})
\label{VV} \\
&=&V(\phi _{0}^{2})+\frac{1}{2}\phi ^{T}\mathcal{T}^{T}H(\phi _{0}^{2})%
\mathcal{T}\phi +\ldots =V(\phi _{0}^{2})+\frac{1}{2}\phi ^{T}H(\phi
_{0}^{2})\phi +\ldots  \notag \\
&=&V(\phi +\phi _{0}^{2}).  \notag
\end{eqnarray}%
As the kinetic term is invariant by itself no modification of the mass
squared matrix will arise from there, apart form the multiplication by $\hat{%
I}$ as a result of the non-Hermitian nature. Thus we may employ the symmetry
to transform the vacuum into the most convenient form for analysis without
altering the physics, such as the eigenvalue spectrum of the mass matrix.

\section{A $\mathcal{CPT}$-symmetric non-Hermitian model with global
SU(2)-symmetry}

Let us now verify the previous general statements for a more concrete
system. We consider the action%
\begin{equation}
\mathcal{I}_{su2}=\int d^{4}x\left[ \sum\limits_{i=1}^{2}\left( \left\vert
\partial _{\mu }\phi _{i}\right\vert ^{2}+m_{i}^{2}\left\vert \phi
_{i}\right\vert ^{2}\right) -\mu ^{2}\left( \phi _{1}^{\dagger }\phi
_{2}-\phi _{2}^{\dagger }\phi _{1}\right) -\frac{g}{4}\left\vert \phi
_{1}\right\vert ^{4}\right] ,  \label{ISU2}
\end{equation}%
where the two complex scalar fields $\phi _{i}=(\phi _{i}^{1},\phi
_{i}^{2})^{T}$, $i=1,2$, are taken to be in the fundamental or spin 1/2
representation of $SU(2)$ and $g,\mu \in \mathbb{R}$ are constants. We allow
here for $m_{i}\in \mathbb{R}$ or $m_{i}\in i\mathbb{R}$, so that $%
m_{i}\rightarrow c_{i}m_{i}$ with $c_{i}=1$ or $c_{i}=-1$, respectively,
takes care of these two possibilities. For simplicity we suppress the
parameters $c_{i}$ until we analyse the physical parameter space in section
3.5. We observe that the action $\mathcal{I}_{su2}$ has the aforementioned
three properties. It is invariant under a global continuous symmetry $\phi
_{j}^{k}\rightarrow \phi _{j}^{k}+\delta \phi _{j}^{k}$ where $\delta \phi
_{j}^{k}=i\alpha _{a}T_{a}^{kl}\phi _{j}^{l}$ with $SU(2)$-Lie algebraic
generators $T_{a}$, is invariant under two discrete antilinear symmetries $%
\mathcal{CPT}_{\pm }:$ $\phi (x_{\mu })\rightarrow \pm \sigma _{3}\phi
^{\ast }(-x_{\mu })$, with $\sigma _{3}$ denoting one of the Pauli spin
matrices, and the potential $V(\phi )$ in (\ref{ISU2}) is evidently not
Hermitian. We note that in the surface term approach \cite{alex2019} the
antilinear symmetries are implemented differently by $\mathcal{PT}$ not
acting on the arguments of the fields.

\subsection{Equivalent Hermitian actions}

More explicitly in components and transformed to the real fields $\varphi
_{j}^{k}$, $\chi _{j}^{k}\in \mathbb{R}$, via $\phi _{j}^{k}=1/\sqrt{2}%
(\varphi _{j}^{k}+i\chi _{j}^{k})$, the action $\mathcal{I}_{su2}$ reads%
\begin{eqnarray}
\mathcal{I}_{su2} &=&\!\!\int d^{4}x\left[ \frac{1}{2}\sum%
\limits_{j,k=1}^{2}\left( \partial _{\mu }\varphi _{j}^{k}\right)
^{2}+\left( \partial _{\mu }\chi _{j}^{k}\right) ^{2}+m_{j}^{2}\left(
\varphi _{j}^{k}\right) ^{2}+m_{j}^{2}\left( \chi _{j}^{k}\right) ^{2}+i2\mu
^{2}\left( \chi _{1}^{k}\varphi _{2}^{k}-\varphi _{1}^{k}\chi
_{2}^{k}\right) \right.  \notag \\
&&\left. -\frac{g}{16}\left[ \left( \varphi _{1}^{1}\right) ^{2}+\left(
\varphi _{1}^{2}\right) ^{2}+\left( \chi _{1}^{1}\right) ^{2}+\left( \chi
_{1}^{2}\right) ^{2}\right] ^{2}\right] .
\end{eqnarray}%
As indicated above, the direct functional variation of this action will lead
to inconsistent equations of motion and we therefore seek a suitable
similarity transformation to resolve this issue. Using the Dyson map%
\begin{equation}
\eta =e^{\frac{\pi }{2}\int d^{3}x\Pi _{2}^{\varphi ,1}(\mathbf{x},t)\varphi
_{2}^{1}(\mathbf{x},t)}e^{\frac{\pi }{2}\int d^{3}x\Pi _{2}^{\varphi ,2}(%
\mathbf{x},t)\varphi _{2}^{2}(\mathbf{x},t)}e^{\frac{\pi }{2}\int d^{3}x\Pi
_{2}^{\chi ,1}(\mathbf{x},t)\chi _{2}^{1}(\mathbf{x},t)}e^{\frac{\pi }{2}%
\int d^{3}x\Pi _{2}^{\chi ,2}(\mathbf{x},t)\chi _{2}^{2}(\mathbf{x},t)},
\label{eta}
\end{equation}%
with canonical momenta $\Pi _{j}^{\varphi ,k}=\partial _{t}\varphi _{j}^{k}$%
, $\Pi _{j}^{\chi ,k}=\partial _{t}\chi _{j}^{k}$ and $\Pi _{j}^{\phi ,k}=$ $%
\partial _{t}\phi _{j}^{k}$, $j,k=1,2$, the adjoint actions of $\eta $ on
the real and complex scalar fields and canonical momenta is computed to%
\begin{eqnarray}
\eta \varphi _{j}^{k}\eta ^{-1} &=&(-i)^{\delta _{2j}}\varphi
_{j}^{k},~~~\eta \chi _{j}^{k}\eta ^{-1}=(-i)^{\delta _{2j}}\chi
_{j}^{k},~~~~~\eta \phi _{j}^{k}\eta ^{-1}=(-i)^{\delta _{2j}}\phi _{j}^{k},
\\
\eta \Pi _{j}^{\varphi ,k}\eta ^{-1} &=&i^{\delta _{2j}}\Pi _{j}^{\varphi
,k},~~~\eta \Pi _{j}^{\chi ,k}\eta ^{-1}=i^{\delta _{2j}}\Pi _{j}^{\chi
,k},~~~~~\eta \Pi _{j}^{\phi ,k}\eta ^{-1}=i^{\delta _{2j}}\Pi _{j}^{\phi
,k}.
\end{eqnarray}%
Thus we can utilize $\eta $ to transform $\mathcal{I}_{su2}$ into a
Hermitian action 
\begin{eqnarray}
&&\hat{{\mathcal{I}}}_{su2}=\eta \mathcal{I}_{su2}\eta ^{-1}=\!\!\int d^{4}x%
\left[ \sum\limits_{j,k=1}^{2}(-1)^{\delta _{2j}}\frac{1}{2}\left[ \left(
\partial _{\mu }\varphi _{j}^{k}\right) ^{2}+\left( \partial _{\mu }\chi
_{j}^{k}\right) ^{2}+m_{j}^{2}\left( \varphi _{j}^{k}\right)
^{2}+m_{j}^{2}\left( \chi _{j}^{k}\right) ^{2}\right] \right.  \notag \\
&&+\left. \mu ^{2}\left( \chi _{1}^{k}\varphi _{2}^{k}-\varphi _{1}^{k}\chi
_{2}^{k}\right) -\frac{g}{16}\left[ \left( \varphi _{1}^{1}\right)
^{2}+\left( \varphi _{1}^{2}\right) ^{2}+\left( \chi _{1}^{1}\right)
^{2}+\left( \chi _{1}^{2}\right) ^{2}\right] ^{2}\right] .  \label{ISU}
\end{eqnarray}%
It is useful to note here for our analysis and especially with regard to the
generalizations to systems with symmetries of higher rank that the action $%
\hat{{\mathcal{I}}}_{su2}$ can also be cast into a more compact form as%
\begin{eqnarray}
\hat{{\mathcal{I}}}_{su2} &=&\int d^{4}x\sum\limits_{i=1}^{2}\partial _{\mu
}\Phi _{i}I\partial ^{\mu }\Phi _{i}+\partial _{\mu }\Psi _{i}I\partial
^{\mu }\Psi _{i}+\frac{1}{2}\Phi _{i}^{T}H_{+}\Phi _{i}+\frac{1}{2}\Psi
_{i}^{T}H_{-}\Psi _{i}  \label{compact} \\
&&-\frac{g}{16}\left( \Phi _{i}^{T}E\Phi _{i}+\Psi _{i}^{T}E\Psi _{i}\right)
^{2},  \notag \\
&=&\int d^{4}x\left[ \partial _{\mu }F\hat{I}\partial ^{\mu }F+\frac{1}{2}%
F^{T}\hat{H}F-\frac{g}{16}\left( F^{T}\hat{E}F\right) ^{2}\right] ,
\label{mc}
\end{eqnarray}%
where we defined the matrices and vectors%
\begin{equation}
H_{\pm }=\left( 
\begin{array}{cc}
m_{1}^{2} & \pm \mu ^{2} \\ 
\pm \mu ^{2} & m_{2}^{2}%
\end{array}%
\right) ,~~I=\left( 
\begin{array}{cc}
1 & 0 \\ 
0 & -1%
\end{array}%
\right) ,~~E=\left( 
\begin{array}{cc}
1 & 0 \\ 
0 & 0%
\end{array}%
\right) ,~~\Phi _{j}=\left( 
\begin{array}{c}
\varphi _{1}^{j} \\ 
\chi _{2}^{j}%
\end{array}%
\right) ,~~\Psi _{j}=\left( 
\begin{array}{c}
\chi _{1}^{j} \\ 
\varphi _{2}^{j}%
\end{array}%
\right) ,
\end{equation}%
$\Phi =(\Phi _{1},\Phi _{2})$, $\Psi =(\Psi _{1},\Psi _{2})$, $F=(\Phi ,\Psi
)=(\varphi _{1}^{1},\chi _{2}^{1},\varphi _{1}^{2},\chi _{2}^{2},\chi
_{1}^{1},\varphi _{2}^{1},\chi _{1}^{2},\varphi _{2}^{2})$, $\limfunc{diag}%
\hat{I}=\{I,I,I,I\}$, $\limfunc{diag}\hat{H}=\{H_{+},H_{+},H_{-},H_{-}\}$, $%
\limfunc{diag}\hat{E}=\{E,E,E,E\}$.

\subsection{$SU(2)$ and $\mathcal{CPT}_{\pm }$-symmetry}

Let us now analyze the model $\hat{{\mathcal{I}}}_{su2}$ in more detail.
First we verify the $SU(2)$-symmetry of the action and its effect on the
different types of fields. Noting that the change in the complex scalar
fields is $\delta \phi _{j}^{k}=i\alpha _{a}T_{a}^{kl}\phi _{j}^{l}$, with
the generators $T_{a}$~of the symmetry transformation taken to be standard
Pauli matrices $\sigma _{a}$, $a=1,2,3$, we directly identify the
infinitessimal changes for the real component fields as%
\begin{eqnarray}
\delta \varphi _{j}^{1} &=&-\alpha _{1}\chi _{j}^{2}+\alpha _{2}\varphi
_{j}^{2}-\alpha _{3}\chi _{j}^{1},\quad \delta \chi _{j}^{1}=\alpha
_{1}\varphi _{j}^{2}+\alpha _{2}\chi _{j}^{2}+\alpha _{3}\varphi _{j}^{1},
\label{t1} \\
\delta \varphi _{j}^{2} &=&-\alpha _{1}\chi _{j}^{1}-\alpha _{2}\varphi
_{j}^{1}+\alpha _{3}\chi _{j}^{2},\quad \delta \chi _{j}^{2}=\alpha
_{1}\varphi _{j}^{1}-\alpha _{2}\chi _{j}^{1}-\alpha _{3}\varphi _{j}^{2}.
\label{t2}
\end{eqnarray}%
It is easily verified that the Hermitian action $\hat{{\mathcal{I}}}_{su2}$
remains invariant under the transformations (\ref{t1}), (\ref{t2}). For the $%
4$ and $8$-component fields the symmetries (\ref{t1}), (\ref{t2}) then
translate into%
\begin{eqnarray}
\delta \Phi &=&-\alpha _{1}\left( \sigma _{1}\otimes \sigma _{3}\right) \Psi
+i\alpha _{2}\left( \sigma _{2}\otimes \mathbb{I}\right) \Phi -\alpha
_{3}\left( \sigma _{3}\otimes \sigma _{3}\right) \Psi , \\
\delta \Psi &=&\alpha _{1}\left( \sigma _{1}\otimes \sigma _{3}\right) \Phi
+i\alpha _{2}\left( \sigma _{2}\otimes \mathbb{I}\right) \Psi +\alpha
_{3}\left( \sigma _{3}\otimes \sigma _{3}\right) \Phi , \\
\delta F &=&i\left[ -\alpha _{1}\left( \sigma _{2}\otimes \sigma _{1}\otimes
\sigma _{3}\right) +\alpha _{2}\left( \mathbb{I}\otimes \sigma _{2}\otimes 
\mathbb{I}\right) -\alpha _{3}\left( \sigma _{2}\otimes \sigma _{3}\otimes
\sigma _{3}\right) \right] F,  \label{dF}
\end{eqnarray}%
with $\otimes $ denoting the standard tensor product. These expressions may
be applied to the action in the forms (\ref{compact}) and (\ref{mc}),
respectively, to verify the $SU(2)$-symmetry.

The antilinear $\mathcal{CPT}_{\pm }$-symmetries manifest themselves as 
\begin{eqnarray}
\mathcal{CPT}_{\pm } &:&~\varphi _{j}^{k}(x_{\mu })\rightarrow \mp
(-1)^{j}\varphi _{j}^{k}(-x_{\mu }),~~\chi _{j}^{k}(x_{\mu })\rightarrow \pm
(-1)^{j}\chi _{j}^{k}(-x_{\mu }), \\
&&\Phi (x_{\mu })\rightarrow \pm \Phi (-x_{\mu }),~~\Psi (x_{\mu
})\rightarrow \mp \Psi (-x_{\mu }), \\
&&F(x_{\mu })\rightarrow \pm \left( \sigma _{3}\otimes \mathbb{I}\otimes 
\mathbb{I}\right) F(-x_{\mu }),
\end{eqnarray}%
which can be verified in (\ref{ISU}), (\ref{compact}) and (\ref{mc}),
respectively.

\subsection{$SU(2)$-symmetry invariant and breaking vacua}

Let us now compute the vacua from (\ref{vac}) with potential as specified in
(\ref{ISU}). We find there are only two types of vacua, that either break or
respect the $SU(2)$-symmetry, 
\begin{eqnarray}
F_{0}^{b} &=&\left( x,-ax,y,-ay,z,az,\pm R,\pm aR\right) ,  \label{v1} \\
F_{0}^{s} &=&\left( 0,0,0,0,0,0,0,0\right) ,  \label{v2}
\end{eqnarray}%
respectively. We introduced the notation $x:=\varphi _{1}^{0,1}$, $y:=$ $%
\varphi _{1}^{0,2}$, $z:=\chi _{1}^{0,1}$, for the vacuum field components
and $a:=\mu ^{2}/m_{2}^{2}$, $R:=\sqrt{r^{2}-(x^{2}+y^{2}+z^{2})}$, $%
r:=4\left( \mu ^{2}+m_{1}^{2}m_{2}^{2}\right) /gm_{2}^{2}$ for convenience.
We note that the defining relation for $R$ can be interpreted as a three
sphere in $\mathbb{R}^{4}$ with center $(0,0,0,0)$ and radius $r$, which is
the geometrical configuration expected from its topological isomorphism with
the $SU(2)$-group manifold. We note that the points $\mu
^{2}=-m_{1}^{2}m_{2}^{2}$ are special as there the three sphere collapses to
a point and the symmetry of the vacuum is restored $F_{0}^{b}\rightarrow
F_{0}^{s}$.

The symmetry properties of the vacua are easily established. Identifying the
generators $T_{a}$~of the symmetry transformation as Pauli matrices, where
we drop the usual factor of $1/2$, we compute the action on the vacuum
states, say $\phi _{j}^{0}=(\phi _{j}^{0,1},\phi _{j}^{0,2})^{T}$ for $j=1,2$%
. We find%
\begin{equation}
T_{1}\phi _{j}^{0}=(\phi _{j}^{0,2},\phi _{j}^{0,1})^{T},\quad ~~T_{2}\phi
_{j}^{0}=(-i\phi _{j}^{0,2},i\phi _{j}^{0,1})^{T},\quad ~~T_{3}\phi
_{j}^{0}=(\phi _{j}^{0,1},-\phi _{j}^{0,2})^{T},  \label{su}
\end{equation}%
so that for non-zero fields the vacuum will always break the symmetry with
respect to the action of $T_{1}$ and $T_{2}$. The action of $T_{3}$ seems to
require only $\phi _{j}^{0,2}=0$, in order to achieve invariance. However,
apart from $F_{0}^{s}$ there is no possible choice for the fields in $%
F_{0}^{b}$ so that $\phi _{j}^{0,1}\neq 0$ in that case.

Let us now make use of the argument in (\ref{VV}) and employ the $SU(2)$%
-symmetry to transform the vacuum $F_{0}^{b}$ into a physically equivalent,
but more manageable one. Choosing two simple target vacua $\check{\phi}%
_{1}^{0}$ and $\check{\phi}_{2}^{0}$, we attempt therefore to simultaneously
solve the two equations%
\begin{eqnarray}
e^{i\alpha _{a}T_{a}}\phi _{1}^{0} &=&\left[ \cos \rho \mathbb{I}+i\sin \rho
(\mathbf{n\cdot \sigma )}\right] \phi _{1}^{0}=\check{\phi}_{1}^{0}=\left( 
\begin{array}{l}
0 \\ 
\pm ir%
\end{array}%
\right) ,  \label{e1} \\
e^{i\alpha _{a}T_{a}}\phi _{2}^{0} &=&\left[ \cos \rho \mathbb{I}+i\sin \rho
(\mathbf{n\cdot \sigma )}\right] \phi _{2}^{0}=\check{\phi}_{2}^{0}=\left( 
\begin{array}{l}
0 \\ 
\pm ar%
\end{array}%
\right) ,  \label{e2}
\end{eqnarray}%
by using the well known formula $e^{i\rho \mathbf{n\cdot \sigma }}=\cos \rho 
\mathbb{I}+i\cos \rho (\mathbf{n\cdot \sigma )}$ with $\rho =\sqrt{\alpha
_{1}^{2}+\alpha _{2}^{2}+\alpha _{3}^{2}}$, $\mathbf{n}=(\alpha _{1},\alpha
_{2},\alpha _{3})/\rho $ and $T_{a}=\sigma _{a}$. The vacuum fields are
parametrized as%
\begin{equation}
\phi _{1}^{0}=\left( 
\begin{array}{l}
\varphi _{1}^{0,1}+i\chi _{1}^{0,1} \\ 
\varphi _{1}^{0,2}+i\chi _{1}^{0,2}%
\end{array}%
\right) =\left( 
\begin{array}{l}
x+iz \\ 
y+iR%
\end{array}%
\right) ,\quad \text{and\quad }\phi _{2}^{0}=\left( 
\begin{array}{l}
\varphi _{2}^{0,1}+i\chi _{2}^{0,1} \\ 
\varphi _{2}^{0,2}+i\chi _{2}^{0,2}%
\end{array}%
\right) =\left( 
\begin{array}{l}
-az+iax \\ 
-aR+iay%
\end{array}%
\right) ,
\end{equation}%
so that the form of the target vacuum is motivated by setting $x=y=z=0$. We
only keep one of the sign in (\ref{v1}) and solve (\ref{e1}), (\ref{e2}) by 
\begin{equation}
x=\frac{r}{\rho }\sin \rho \alpha _{1},~~~~y=-\frac{r}{\rho }\sin \rho
\alpha _{3},~~~~~z=-\frac{r}{\rho }\sin \rho \alpha _{2},
\end{equation}%
so that $R=r\cos \rho $. For the vacuum $F_{0}^{b}$ this translates with (%
\ref{dF}) into%
\begin{equation}
\mathcal{T}F_{0}^{b}=\check{F}_{0}^{b},  \label{TF}
\end{equation}%
where%
\begin{eqnarray}
\mathcal{T} &=&\cos (\rho )\mathbb{I}_{8}-i\frac{\sin (\rho )}{\rho }\left[
\alpha _{1}\left( \sigma _{2}\otimes \sigma _{1}\otimes \sigma _{3}\right)
-\alpha _{2}\left( \mathbb{I}\otimes \sigma _{2}\otimes \mathbb{I}\right)
+\alpha _{3}\left( \sigma _{2}\otimes \sigma _{3}\otimes \sigma _{3}\right) %
\right] ,~~~~~\  \\
\check{F}_{0}^{b} &=&\left( 0,0,0,0,0,0,\pm r,\pm ar\right) .
\end{eqnarray}%
We note that $\det \mathcal{T}=1$ and as required $\mathcal{T}^{T}=\mathcal{T%
}^{-1}$. Evidently $\check{F}_{0}^{b}$ is of a more convenient form of the
vacuum than $F_{0}^{b}$ and we shall therefore use it from here on.

\subsection{Mass squared matrices and null vectors}

Next we use the different vacua and expand the potentials around them to
determine the mass squared matrix according to the definition in (\ref{Gold}%
). Expanding first around the $SU(2)$-symmetric vacuum $F_{0}^{s}$ we find
the mass squared matrix%
\begin{equation}
M_{s}^{2}=\left( 
\begin{array}{cccccccc}
-m_{1}^{2} & \mu ^{2} & 0 & 0 & 0 & 0 & 0 & 0 \\ 
-\mu ^{2} & -m_{2}^{2} & 0 & 0 & 0 & 0 & 0 & 0 \\ 
0 & 0 & -m_{1}^{2} & \mu ^{2} & 0 & 0 & 0 & 0 \\ 
0 & 0 & -\mu ^{2} & -m_{2}^{2} & 0 & 0 & 0 & 0 \\ 
0 & 0 & 0 & 0 & -m_{1}^{2} & -\mu ^{2} & 0 & 0 \\ 
0 & 0 & 0 & 0 & \mu ^{2} & -m_{2}^{2} & 0 & 0 \\ 
0 & 0 & 0 & 0 & 0 & 0 & -m_{1}^{2} & -\mu ^{2} \\ 
0 & 0 & 0 & 0 & 0 & 0 & \mu ^{2} & -m_{2}^{2}%
\end{array}%
\right) ,
\end{equation}%
with two fourfold degenerate eigenvalues%
\begin{equation}
\lambda _{\pm }^{s}=-\frac{1}{2}\left( m_{1}^{2}+m_{2}^{2}\pm \sqrt{%
(m_{1}^{2}-m_{2}^{2})^{2}-4\mu ^{4}}\right) .
\end{equation}%
As expected from (\ref{Gold}) there are no Goldstone bosons emerging in this 
$SU(2)$-invariant case.

Expanding instead around the $SU(2)$-symmetry breaking vacuum $F_{0}^{b}$,
we obtain the mass squared matrix 
\begin{equation}
M_{b}^{2}=\left( 
\begin{array}{cccccccc}
\frac{g(\varphi _{1}^{1})^{2}}{2}+\frac{\mu ^{4}}{m_{2}^{2}} & \mu ^{2} & 
\frac{g\varphi _{1}^{1}\varphi _{1}^{2}}{2} & 0 & \frac{g\varphi
_{1}^{1}\chi _{1}^{1}}{2} & 0 & -\frac{\varphi _{1}^{1}gR}{2} & 0 \\ 
-\mu ^{2} & -m_{2}^{2} & 0 & 0 & 0 & 0 & 0 & 0 \\ 
\frac{g\varphi _{1}^{1}\varphi _{1}^{2}}{2} & 0 & \frac{g(\varphi
_{1}^{2})^{2}}{2}+\frac{\mu ^{4}}{m_{2}^{2}} & \mu ^{2} & \frac{g\varphi
_{1}^{2}\chi _{1}^{1}}{2} & 0 & -\frac{\varphi _{1}^{2}gR}{2} & 0 \\ 
0 & 0 & -\mu ^{2} & -m_{2}^{2} & 0 & 0 & 0 & 0 \\ 
\frac{g\varphi _{1}^{1}\chi _{1}^{1}}{2} & 0 & \frac{g\varphi _{1}^{2}\chi
_{1}^{1}}{2} & 0 & \frac{g(\varphi _{1}^{2})^{2}}{2}+\frac{\mu ^{4}}{%
m_{2}^{2}} & -\mu ^{2} & -\frac{\chi _{1}^{1}gR}{2} & 0 \\ 
0 & 0 & 0 & 0 & \mu ^{2} & -m_{2}^{2} & 0 & 0 \\ 
-\frac{\varphi _{1}^{1}gR}{2} & 0 & -\frac{\varphi _{1}^{2}gR}{2} & 0 & -%
\frac{\chi _{1}^{1}gR}{2} & 0 & \frac{g^{2}R^{2}}{2}+\frac{\mu ^{4}}{%
m_{2}^{2}} & -\mu ^{2} \\ 
0 & 0 & 0 & 0 & 0 & 0 & \mu ^{2} & -m_{2}^{2}%
\end{array}%
\right) .
\end{equation}%
The expansion around $\check{F}_{0}^{b}$ yields the same matrix with $%
\varphi _{1}^{1}=\chi _{1}^{1}=\varphi _{1}^{2}=0$. As expected from (\ref%
{VV}) and (\ref{TF}), both matrices share the same field independent
eigenvalues, that is two different ones each with a threefold degeneracy and
two eigenvalues that may give rise to an exceptional point%
\begin{equation}
\lambda _{1,2,3}^{b}=0,~~~\lambda _{4,5,6}^{b}=\frac{\mu ^{4}}{m_{2}^{2}}%
-m_{2}^{2},~~~\lambda _{\pm }^{b}=K\pm \sqrt{K^{2}+2L}.
\end{equation}%
For convenience we defined here $K:=3\mu
^{4}/2m_{2}^{2}+m_{1}^{2}-m_{2}^{2}/2$ and $L:=\mu ^{4}+m_{1}^{2}m_{2}^{2}$.
We confirm the expectation from Goldstone's theorem to find three massless
Goldstone bosons in the symmetry breaking sector, since none of the three $%
SU(2)$-generators leaves the vacuum $F_{0}^{b}$ invariant.

According to the relation (\ref{Gold}) we may compute the corresponding null
vectors directly from the $SU(2)$-symmetry transformation. When applying the
infinitessimal changes for the component fields (\ref{t1}) and (\ref{t2}) to
the vacuum $F_{0}^{b}$, we obtain the vectors%
\begin{eqnarray}
\nu _{1}^{0} &=&\frac{1}{\sqrt{N}}\left\{ R,-aR,-\chi _{1}^{1},\frac{\mu
^{2}\chi _{1}^{1}}{m_{2}^{2}},\varphi _{1}^{2},\frac{\mu ^{2}\varphi _{1}^{2}%
}{m_{2}^{2}},\varphi _{1}^{1},\frac{\mu ^{2}\varphi _{1}^{1}}{m_{2}^{2}}%
\right\} , \\
\nu _{2}^{0} &=&\frac{1}{\sqrt{N}}\left\{ \varphi _{1}^{2},-\frac{\mu
^{2}\varphi _{1}^{2}}{m_{2}^{2}},-\varphi _{1}^{1},\frac{\mu ^{2}\varphi
_{1}^{1}}{m_{2}^{2}},-R,-aR,-\chi _{1}^{1},-\frac{\mu ^{2}\chi _{1}^{1}}{%
m_{2}^{2}}\right\} , \\
\nu _{30} &=&\frac{1}{\sqrt{N}}\left\{ -\chi _{1}^{1},\frac{\mu ^{2}\chi
_{1}^{1}}{m_{2}^{2}},-R,aR,\varphi _{1}^{1},\frac{\mu ^{2}\varphi _{1}^{1}}{%
m_{2}^{2}},-\varphi _{1}^{2},-\frac{\mu ^{2}\varphi _{1}^{2}}{m_{2}^{2}}%
\right\} ,
\end{eqnarray}%
with $N:=-4L\lambda _{4,5,6}^{b}/gm_{2}^{4}$. These vectors have been
normalized with regard to the $\mathcal{CPT}$-inner product $\left\langle
x\right. \left\vert Iy\right\rangle $. We verify that the $\nu _{i}^{0}$, $%
i=1,2,3$, are indeed null vectors of $M_{b}^{2}$. Furthermore, we observe
from the normalization constant that at the zero exceptional points, i.e.
for $\mu ^{4}=m_{2}^{4}$ when $\lambda _{4,5,6}^{b}=0$ and $\mu
^{4}=-m_{1}^{2}m_{2}^{2}$ when $\lambda _{-}^{b}=0$, these vectors are not
defined. We may ignore the case\ $\lambda _{-}^{b}=0$ in what follows as in
this case the $SU(2)$-symmetry is restored with $\check{F}%
_{0}^{b}\rightarrow F_{0}^{s}$.

\subsection{Physical regions}

We will now analyse the parameter space of the system and identify the
physical regions based on a meaningful mass squared matrix. To cover all
possible cases we are setting therefore in all expressions $m_{i}^{2}$ $%
\rightarrow c_{i}m_{i}^{2}$.For the model expanded around the broken vacuum
the physical regions are then determined by $\lambda _{\pm }^{b}\geq 0$, $%
\lambda _{4,5,6}^{b}\geq 0$ corresponding to the four inequalities 
\begin{equation}
K\geq 0,~~~~L\leq 0,~~~K^{2}+2L\geq 0,~~~~c_{2}\mu ^{4}\geq c_{2}m_{2}^{4},
\end{equation}%
for the four cases $c_{1}=\pm 1$, $c_{2}=\pm 1$. All constraints can be
expressed as functions of the two ratios $(\mu
^{4}/m_{1}^{4},m_{2}^{2}/m_{1}^{2})$. We find that no solutions exists for $%
c_{1}=$ $c_{2}$, apart from setting $\mu =m_{2}=0$, so that in these two
case the model is unphysical. The physical regions for the remaining two
cases $c_{1}=-$ $c_{2}=\pm 1$ are depicted in figure \ref{Fig1}.

\FIGURE{ \epsfig{file=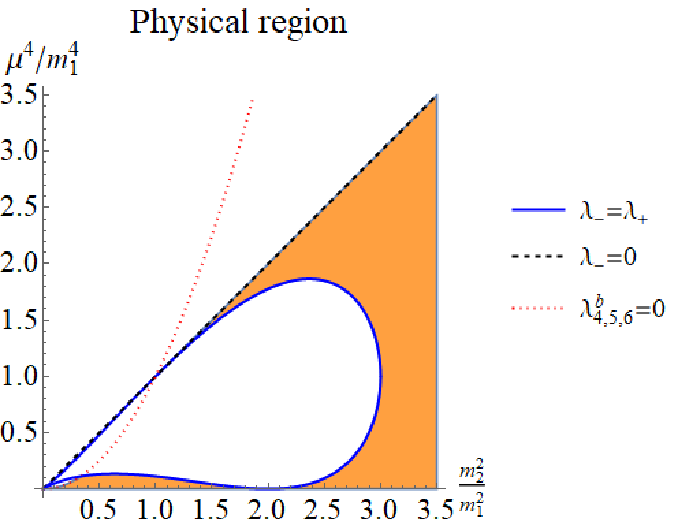, height=6.5cm} \epsfig{file=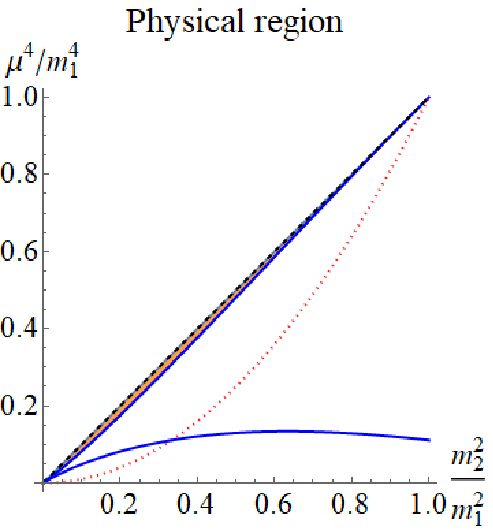, height=6.5cm}
\caption{Physical regions in parameter space bounded by exceptional and zero exceptional points as function of $(\mu^{4}/m_{1}^{4},m_{2}^{2}/m_{1}^{2})$ 
for the theory expanded around the SU(2)-symmetry breaking vacuum. Left panel for $c_1=-c_2=1$ and right panel for $c_1=-c_2=-1$.}
       \label{Fig1}}

The two different cases depicted in figure \ref{Fig1} do not have any
physical regions that intersect. The case $c_{1}=-$ $c_{2}=1$ was also
analysed within the \textit{surface term approach in }\cite{alex2019} and
our results appear to match exactly. The case $c_{1}=-$ $c_{2}=-1$ was not
dealt with in \cite{alex2019}, but as depicted in figure \ref{Fig1}, it also
contains a well defined small physical region. We note that for our model
with two complex scalar fields the physical regions have no boundary
corresponding to singularities, which appears to be a feature only occurring
for the theories with more complex scalar fields, see \cite%
{fring2019goldstone}.

Finally in figure \ref{Fig2} we also depict the physical regions for the
model expanded around the $SU(2)$-invariant vacuum.

\FIGURE{ \epsfig{file=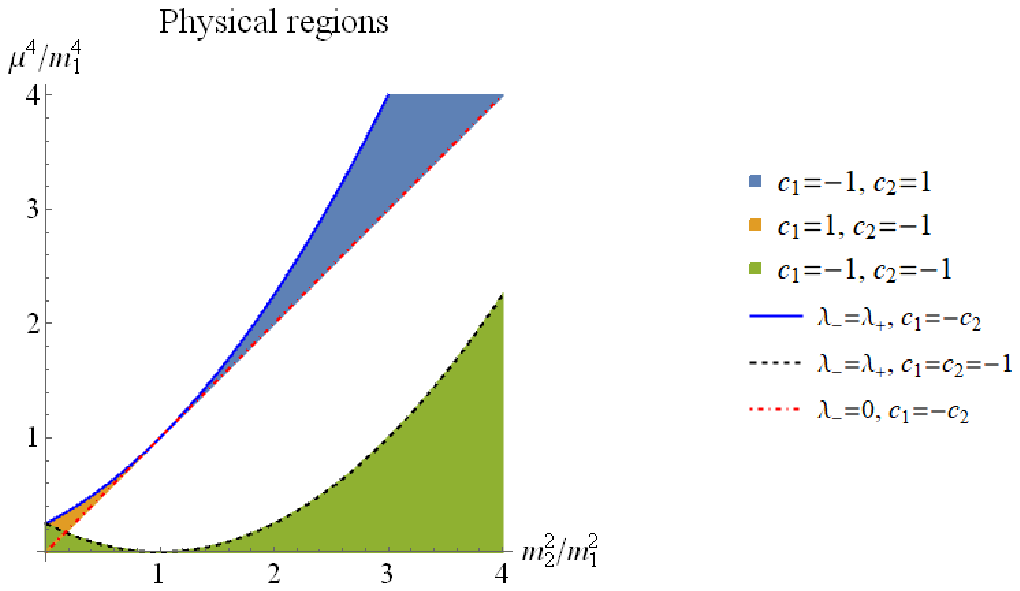, width=14.5cm}
\caption{Physical regions in parameter space bounded by exceptional and zero exceptional points as function of $(\mu
^{4}/m_{1}^{4},m_{2}^{2}/m_{1}^{2})$ for the theory expanded around the SU(2)-symmetry invariant vacuum. }
        \label{Fig2}}

Here only the case $c_{1}=$ $c_{2}=1$ does not contain a physical region
apart from $\mu =m_{2}=0$. The three different cases depicted in figure \ref%
{Fig2} do not have any physical regions that intersect, apart from the small
region near the origin. Comparing figures \ref{Fig1} and \ref{Fig2} we also
notice that cases with equal choices for the $c_{i}$ do not share physical
regions. This implies that for any particular physical model the breaking of
the $SU(2)$-symmetry leads to an unphysical model and in reverse also that
some unphysical models become physical when the $SU(2)$-symmetry is broken.

\subsection{The Goldstone bosons in the $\mathcal{PT}$-symmetric regime}

We may now compute the Goldstone bosons in terms of the original fields in a
similar fashion as discussed in \cite{fring2019goldstone}. Defining for this
purpose the remaining right eigenvectors $v_{i}$, $i=4,\ldots ,8$, and a
matrix $U$ containing all of them\ as column vectors as%
\begin{equation}
M_{b}^{2}v_{i}=\lambda _{i}^{b}v_{i},\qquad
U:=(v_{1},v_{4},v_{2},v_{5},v_{3},v_{6},v_{-},v_{+}),\quad i=1,\ldots ,6,\pm
,  \label{FMU}
\end{equation}%
we diagonalize the mass squared matrix by means of the similarity
transformation $U^{-1}M_{b}^{2}U=D$ with $\limfunc{diag}D=(\lambda
_{1}^{b},\lambda _{4}^{b},\lambda _{2}^{b},\lambda _{5}^{b},\lambda
_{3}^{b},\lambda _{6}^{b},\lambda _{-}^{b},\lambda
_{+}^{b})=(m_{1}^{2},\ldots ,m_{8}^{2})$. For $\mu ^{4}\neq m_{2}^{4}$ and $%
K^{2}\neq -2L$, that are the zero and standard exceptional points, we define
the fields $\psi _{i}$ with masses $m_{i}$ by re-writing the squared mass
term as%
\begin{equation}
F^{T}M_{b}^{2}F=\sum\nolimits_{k=1}^{8}m_{k}^{2}\psi
_{k}^{2}=\sum\nolimits_{k=1}^{8}m_{k}^{2}(F^{T}IU)_{k}(U^{-1}F)_{k}.
\end{equation}%
Hence, the three Goldstone fields are identified as%
\begin{equation}
\psi _{\ell }^{\text{Gb}}:=\sqrt{(F^{T}IU)_{\ell }(U^{-1}F)_{\ell }},~~\
\ell =1,3,5.  \label{FB}
\end{equation}%
Setting in $M_{b}^{2}$ the fields $\chi _{1}^{0,1}$, $\varphi _{1}^{0,1}$, $%
\varphi _{1}^{0,2}$ to zero we compute%
\begin{equation}
U=\left( 
\begin{array}{ccccc}
H_{-} & 0 & 0 & 0 & 0 \\ 
0 & H_{-} & 0 & 0 & 0 \\ 
0 & 0 & H_{+} & 0 & 0 \\ 
0 & 0 & 0 & \lambda _{-}^{b}+m_{2}^{2} & \lambda _{+}^{b}+m_{2}^{2} \\ 
0 & 0 & 0 & \mu ^{2} & \mu ^{2}%
\end{array}%
\right) ,  \label{U}
\end{equation}%
with $\det U=2\mu ^{2}(\mu ^{4}-m_{2}^{4})^{3}\sqrt{K^{2}+2L}$, so that the
explicit form of the Goldstone boson fields in the original fields result to%
\begin{equation}
\psi _{1}^{\text{Gb}}=\frac{\mu ^{2}\varphi _{2}^{1}-m_{2}^{2}\chi _{1}^{1}}{%
\sqrt{m_{2}^{4}-\mu ^{4}}},\qquad \psi _{3}^{\text{Gb}}=\frac{%
m_{2}^{2}\varphi _{1}^{2}+\mu ^{2}\chi _{2}^{2}}{\sqrt{m_{2}^{4}-\mu ^{4}}}%
,\qquad \psi _{5}^{\text{Gb}}=\frac{m_{2}^{2}\varphi _{1}^{1}+\mu ^{2}\chi
_{2}^{1}}{\sqrt{m_{2}^{4}-\mu ^{4}}}.  \label{FGex}
\end{equation}%
As $U$ is not invertible at the exceptional points for $\mu ^{4}=m_{2}^{4}$
and $K^{2}=-2L$, we need to treat these cases separately. We note that these
expressions differ from those obtained in \cite{alex2019}.

\subsection{The Goldstone bosons at the exceptional point}

At the standard exceptional point, i.e. when $K^{2}=-2L$ and hence $\lambda
_{+}^{b}=\lambda _{-}^{b}$, the two eigenvectors $v_{-}$ and $v_{+}$
coalesce so that the matrix $U$ is no longer invertible and the Goldstone
boson fields may take on a different form as found in \cite%
{fring2019goldstone}. Instead of diagonalising the mass squared matrix we
can convert it into Jordan normal form by means of a similarity
transformation. Making $m_{1}$ the dependent variable, the exceptional point
occurs when $m_{1}^{2}=\pm \mu ^{2}-m_{2}^{2}/2-3\mu ^{4}/2m_{2}^{2}$ so
that the Jordan normal form becomes%
\begin{equation}
\limfunc{diag}D_{e}=(0,\lambda _{e}^{b},0,\lambda _{e}^{b},0,\lambda
_{e}^{b},\Lambda ),\text{~~~}\lambda _{e}^{b}=\frac{\mu ^{4}}{m_{2}^{2}}%
-m_{2}^{2},~~~\Lambda =\left( 
\begin{array}{ll}
\pm \mu ^{2}-m_{2}^{2} & \pm (\alpha -\beta )\mu ^{2} \\ 
0 & \pm \mu ^{2}-m_{2}^{2}%
\end{array}%
\right) ,
\end{equation}%
which can be obtained from the similarity transformation $%
U_{e}^{-1}M_{e}^{2}U_{e}=D_{e}$ with $U_{e}$ equalling $U$ with the lower
right block replaced by%
\begin{equation}
\left( 
\begin{array}{ll}
1 & \alpha  \\ 
1 & \beta 
\end{array}%
\right) .
\end{equation}%
We compute now $\det U=(\alpha -\beta )(\mu ^{4}-m_{2}^{4})^{3}$. Defining
the Goldstone boson fields by the same formal expression as in (\ref{FB}),
but with $U$ replaced by $U_{e}$, we obtain at the exceptional point the
same expressions as in (\ref{FB}). It is worth noting that the two
degenerate fields take on the form 
\begin{equation}
\psi _{+,e}=\psi _{-,e}=\frac{\sqrt{(\varphi _{2}^{2}-\chi _{1}^{2})(\alpha
\varphi _{2}^{2}-\beta \chi _{1}^{2})}}{\sqrt{\beta -\alpha }}.
\end{equation}%
We note that it is by far not obvious that the Goldstone boson fields
acquire the same form in the $\mathcal{PT}$-symmetric regime as at the
exceptional point. This is more a coincidence due to the special nature of
the mass matrix rather than a general feature. When considering models with
more than two scalar fields this no longer holds even for the Abelian case
as observed in \cite{fring2019goldstone}. In \cite{alex2019} this regime was
not analysed separately.

\section{Conclusions and outlook}

Using a pseudo-Hermitian approach to treat non-Hermitian quantum field
theories we found that the Goldstone theorem also holds when the global
symmetry group is non-Abelian. The explicit form for the Goldstone boson in
the $\mathcal{PT}$-symmetric regime and at the standard exceptional points
can be found explicitly, although using different diagonalisation procedures
for the mass squared matrix. At the zero exceptional point the Goldstone
boson can not be identified. When the analysis of our model overlaps with
the one carried out in \cite{alex2019} employing the surface term approach,
the physical regions coincide exactly. However, the explicit forms of the
Goldstone bosons are different.

There are some obvious further extensions to these investigation, that would
be interesting to carry out, such as the treatment of models with different
Lie symmetry groups and the augmentation of the amount of complex scalar
fields. Most interesting, with regard to the comparison with the surface
term approach, is the investigation of the Higgs mechanism within the
presented framework as that aspect will produce more features and
predictions that are clearly distinct in the two approaches \cite{AFTTprep}.

Furthermore, it would be very interesting to establish a closer link between
studies carried on non-Hermitian systems in 1+1 dimensions. In principle,
the Goldstone theorem does not apply for dimension $d\leq 2$ as in those
settings the breaking of continuous symmetries inevitably leads to infrared
divergent correlation functions. However, in \cite{jacobsen2003} it was
argued that the Mermin-Wagner theorem no longer applies for the continuous $%
SO(N)$-symmetry with $N<2$ as it cannot be realized as unitary operations on
a vector fields. This feature was exploited in \cite{jacobsen2003} to
identify a Goldstone phase for a non-Hermitian system..

\noindent \textbf{Acknowledgments:} We would like to thank Peter Millington
for interesting discussions and Hubert Saleur for pointing out reference 
\cite{jacobsen2003} to us.

\newif\ifabfull\abfulltrue


\end{document}